\documentclass{myart}

\usepackage[T1]{fontenc}

\usepackage{graphicx}					
\usepackage{todonotes}

\usepackage{pgf}			
\usepackage{tikz}						
\newcommand{\hide}[1]{}
\newcommand{\code}[1]{\textsf{\small #1}}

\begin{document}

\title{How Developers Extract Functions:\\ An Experiment%
\thanks{This research was supported by the ISRAEL SCIENCE FOUNDATION (grant no.\ 832/18).}}

\author{\large\rule{0pt}{13pt}
Alexey Braver \hspace*{15mm} Dror G. Feitelson\\[5pt]
\normalsize Department of Computer Science\\
\normalsize The Hebrew University of Jerusalem, Israel}

\maketitle

\begin{abstract}
    Creating functions is at the center of writing computer programs.
    But there has been little empirical research on how this is done and what are the considerations that developers use.
    We design an experiment in which we can compare the decisions made by multiple developers under exactly the same conditions.
    The experiment is based on taking existing production code, ``flattening'' it into a single monolithic function, and then charging developers with the task of refactoring it to achieve a better design by extracting functions.
    The results indicate that developers tend to extract functions based on structural cues, such as \code{if} or \code{try} blocks.
    And while there are significant correlations between the refactorings performed by different developers, there are also significant differences in the magnitude of refactoring done.
    For example, the number of functions that were extracted was between 3 and 10, and the amount of code extracted into functions ranged from 37\% to 95\%.
\end{abstract}


\section{Introduction}
\label{sect:Introduction}

Almost every field in our life now includes computers and software.
Many resources are continuously invested in developing new algorithms and new functionalities to support customer needs.
However, the internal design of a product is no less important.
In particular, the design can have a significant impact on the continued development and maintenance of the software.
A good design can save a lot of effort for the developers, and as a result decrease the development time of the product and facilitate faster response to evolving needs.

It is common to divide software design into two main parts:
\begin{itemize}
   \item \emph{High-Level Design:} The overall design of the product showing how the business requirements are satisfied.
   This consists of a description of the architecture, as designed by the software architect.
   \item \emph{Low-Level Design:} The detailed design of the product based on refining the high-level design.
   Mostly, this is created on the fly by the system's developers.
\end{itemize}

A core activity in low-level design is to define functions.
Functions are the basic elements for structuring code.
Creating functions is one of the first things that one learns when learning to program.
Famous coding guides like Bob Martin's \textit{Clean Code} \cite{martin:clean} and Steve McConnell's \textit{Code Complete} \cite{mcconnell:cc} devote full chapters to writing functions.
Martin, for example, opines \cite[pg.\ 34]{martin:clean},
\begin{quote}
    The first rule of functions is that they should be small. The second rule of functions is that \emph{they should be smaller than that}. This is not an assertion that I can justify. I can’t provide any references to research that shows that very small functions are better. What I can tell you is that for nearly four decades I have written functions of all different sizes. I’ve written several nasty 3,000-line abominations. I’ve written scads of functions in the 100 to 300 line range. And I’ve written functions that were 20 to 30 lines long. What this experience has taught me, through long trial and error, is that functions should be very small.
\end{quote}
And indeed, there has been relatively little research about functions and how they are defined.
In particular, we have found no controlled experiments on this topic.
A possible reason is that developers enjoy considerable freedom during low level design.
Due to this freedom, it is impractical to conduct an experiment just by giving a coding task to the participants, and asking them to implement it, because it can be very difficult to compare the results.

Our solution to this problem is to base the experiment on refactoring a given code-base, rather than asking participants to produce new code.
And while we do not answer Martin's explicit question concerning the best length for functions, we do take a first step towards showing what sizes developers prefer and how function creation can be studied experimentally.
Our main contributions in this work are:
\begin{itemize}
    \item To devise a methodology for studying function creation by diverse experimental subjects, at least in the context of refactoring.
    \item To demonstrate that code structure provides cues for function extraction.
    \item To find that developers tend to use short functions even if this is not their explicit goal.
\end{itemize}

\section{Background and Related Work}
\label{sect:related}

\subsection{Refactoring and Function Extraction}
\label{sect:refactor} 

Product requirements always change over time, causing the software to evolve \cite{lehman80,lehman96}.
This may also require adaptions in the software design, namely refactoring \cite{fowler:refactor}.
And one of the basic and most widely used actions performed during refactoring is function extraction, that is turning a block of code into a separate function \cite{murphyh12,tsantalis11,tsantalis18,hora20}.
This may be done for a variety of reasons \cite{liuliu16}:
\begin{itemize}
    \item \emph{Reuse:}\label{ext:reuse} 
    An often cited reason for creating functions is to reuse the code.
    In the context of refactoring this may happen if we identify repetitions in the code (code clones) \cite{li09,tairas12,jbara14a}.
    
    \item \emph{Readability:}\label{ext:readability} 
    Extracting a function and naming it with an informative name makes the code more understandable, especially if it is given an informative name \cite{lammel02}.

    \item \emph{Understandability:}\label{ext:big-functions}  
    A special case of improving code is breaking very long functions \cite{haas16}.
    By extracting sub-computations (together with the local variables used only in them) the function can change into a short sequence of calls which delineate its logic, with local variables only to connect these parts \cite{tsantalis11}.

    \item \emph{Reflect flow:}
    Another aspect of readability worth mentioning is reflecting the program's flow:
    different paths are placed in different functions.
    In particular, in \textit{Clean Code} Martin suggests that try-catch blocks should be extracted, thereby separating functionality from error handling \cite[p.\ 46]{martin:clean}.

    \item \emph{Better testing:}\label{ext:testing} 
    Long functions with many responsibilities are also hard to test: an exponential number of tests may be needed to exercise all the combinations of behaviors, and testing deeply nested code requires a precise identification of the conditions to reach it.
    McCabe suggested that function with a cyclomatic complexity above 10 should be refactored \cite{mccabe76}.

    \item \emph{Expose API:}\label{ext:API} 
    When we extract functions and define them public, we expose their usage outside the class.
    This not only allows these functions to be used outside the class, but also helps to define the functionality of the class.
    Creating a function also provides the opportunity to extend or override behavior in sub-classes.
\end{itemize}

There has been extensive research on tools to perform function extraction semi-automatically.
These are usually based on the identification of \emph{program slices} --- subprograms responsible for the computation of the value of some variable \cite{weiser84}.
For example, Maruyama proposed a tool where the user starts by identifying a variable of interest \cite{maruyama01}.
The tool then finds program slices that affect this variable, and proposes methods that extract these slices.
Komondoor and Horwitz identify conditions under which semantic equivalence holds after extracting some code, and develop algorithms to handle problematic cases with non-contiguous code and non-local jumps (\code{goto}, \code{continue}, and \code{break}) \cite{komondoor00,komondoor03}.
Ettinger and Verbaere proposed a tool to extract separate concerns into aspects \cite{ettinger04}, and Ettinger later formalized slicing-based tools \cite{ettinger07}.
Tsantalis and Chatzigeorgiou suggested using object state slices, and additional program analysis techniques \cite{tsantalis11}.
Haas and Hummel propose to score candidate extracted functions by how much they reduce the complexity of the original long function \cite{haas16}.
This is done based on length, nesting level, and number of required parameters.

Data collected on using refactoring tools indicates that function extraction is one of the most common types of refactoring.
Half of all refactoring done with JDeodorant were function extractions \cite{tsantalis18}.
An analysis of more than 400 thousand refactorings by Hora and Robbes found that 17\% were method extraction \cite{hora20}.
But we know of no research about how developers actually perform function extraction, especially when not using tools.

\subsection{Function Length and Code Quality}
\label{sect:func-lit} 

Refactoring reflects a perception of technical debt \cite{allman12,tom13}: code that is not as clean as it could be, perhaps due to time pressure when it was written.
Function extraction, like other types of refactoring, is expected to improve the quality of the code and make it easier to work with \cite{zabardast20}.
A possible improvement is to make functions shorter, but there has been very little empirical research on how short they should be.
The only study we know of which specifies a concrete optimal function size is Banker et al.\ \cite{banker93}.
This 30 year old study analyzed the maintenance of large Cobol projects, and found the optimal procedure size to be 44 statements.
But Fenton and Neil argue against the ``Goldilocks conjecture'', that there is an ideal size that is not too small and not too large \cite{fenton99}.

Several studies have focused on identifying long methods using metrics for size, cohesion, and coupling \cite{yoshida16,charalampidou15,charalampidou18}.
One of the reasons function extraction is considered beneficial is that shorter functions will be less complex.
Intuitively it is plausible that longer code is harder to understand, but does that mean that if we refactor it into separate functions it will be more understandable?
Several studies have found that code complexity is indeed related to scope \cite{banker93}, and some have claimed that this is such a strong correlation that length is the only metric that is needed \cite{herraiz:loc,gil17}.
Landman et al., in contradistinction, claim that the correlation between code complexity and lines of code is not as high as previously thought \cite{landman16}.
Their view is that high correlations are the result of aggregation, and if individual functions are considered then complexity metrics provide additional information to length.
This echoes earlier work by Gill and Kemerer, who propose to use ``complexity density'', namely to normalize complexity metrics by length \cite{gill91}.
Similar disagreements have been voiced about the size of modules and classes \cite{fenton99}.
El Emam et al.\ also argue against the claim that small components are necessarily beneficial, and fail to find evidence for a threshold beyond which large classes cause more defects \cite{elemam02}.

As noted above, Martin claims in his book \textit{Clean Code} that functions should be as small as possible \cite{martin:clean}.
A survey of practitioners indicates that there is wide acceptance of this suggestion \cite{ljung22}.
Martin goes on to claim that the block of code within \code{if} statements, \code{else} statements, \code{while} statements, and so on should be one line long --- and probably this line should be a function call with a descriptive name that adds documentary value.
This approach emphasizes explanation and documentation of the code as the reasons for function extraction.
A similar stand is taken by Clausen, who uses the rule that functions should be no more than 5 lines of code as the title of his book on refactoring \cite{clausen:5lines}.
In our research we will try to check if these practices are applied by developers.

\section{Research Questions}
\label{sect:Research Questions}

We are motivated by very basic questions in programming, and specifically by the question of function length.
For example, how does function length interact with testing, with the propensity for having defects, and with understanding the whole system?
These questions are interesting and important, but complicated, because myriad factors and considerations may affect the design of code.
In addition, the answer may be ill-defined, as it may depend on the specific developers involved in writing and maintaining the code, who could be different from each other and from other developers.

To make some progress we therefore start with much more limited and concrete questions.
We focus on function extraction rather than the whole issue of writing functions.
Within this context we ask
\begin{enumerate}
    \item How do developers extract functions?
    \item Do developers agree regarding how to refactor code by extracting functions?
    \item What are the considerations that developers apply when deciding about functions?
    \item Do developers conform with the recommendation in \textit{Clean Code} to use very short functions?
\end{enumerate}
To answer these questions, we conduct an experiment in which developers are asked to refactor monolithic code.
We then ask them to answer a short survey about their considerations.

\section{Methodology}
\label{sect:Methods}

Our experiment was based on real production code.
This code was modified to make it monolithic by inlining all the functions.
The experiment participants then refactored it, and we analyzed the resulting codes.

\subsection{Code Selection}
\label{sect:Code}

Our goal was to see how different developers create functions.
This is a non-trivial experiment, because if we just give them a programming assignment that includes enough logic to get different designs, they can come up with very different designs that are hard to compare.
As an alternative we decided to start from a given code-base and just ask them to refactor.

To make the experiment as realistic and relevant as possible we decided to use real production code as the basis.
The main consideration was that it be a general utility that can be understood with reasonable effort.
It also had to be long enough, and have enough logic, so that it would be reasonable to expect some variations in the results.
In other words, we needed to avoid code that could be arranged in only one reasonable way.

After we found a suitable code, we performed a pilot and sent the experiment to 4 developers.
The results we received indicated that the experiment works.
The code we chose is from file \code{adapters.py} in the \code{Requests} http Python library%
\footnote{https://github.com/psf/requests/blob/master/requests/adapters.py}.
This module allows to send HTTP/1.1 requests easily.
The code was originally 534 lines long including comments and blank lines.
Excluding blank lines leaves 434 lines, and also excluding comments leaves only 286 lines of actual code.

\subsection{Code Flattening}
\label{sect:flatteningt}

We call the procedure of converting the code from many functions to one function ``flattening''.
The code was flattened manually in the following way:
\begin{enumerate}
    \item One function was chosen to be the main function (\code{send}).
    This function was the biggest function in the original code as well.
    It included the main logic and calls to all the smaller functions.

    \item Every function which was only defined, but not called internally, was removed.
    These were functions which were only called from other classes.
    We decided to remove them since it would be unreasonable to include unused code in our monolithic function.

    \item \label{Inlined functions}The remaining functions were inlined by replacing calls to them with the function's code.
    The names of each function's parameters were replaced with the names of the arguments the function was called with. 
    In one case we also needed to change a function name that was the same as that of a function from a different module (changed \code{send} to \code{send\_the\_request}).
 
    \item All unrelated classes and imports were removed.
    There were some additional classes which were only defined but not used.
    We remove them to prevent the code from being too long and cluttered.

    \item All the comments in the original code which were added by its developers to describe some flow in the code, and the main function definition, were left as they are.
    We wanted the participants to experience the code as close to the original as possible.
    But the header comments of the inlined functions had to be removed, since there was no suitable place for them.

    \item All calls to third-party functions remained as they are.
    We wanted the participants to refactor the code in a specific context, and not in such large context as all the Python language.
    So we inlined only the project's functions, and left calls to functions from other sources.
\end{enumerate}
All these changes reduced the total length of the code to 298 lines including blank lines and comments, and 208 lines of actual code.
This is the code the participants received to refactor.

\subsection{Experiment Execution}
\label{sec:Experimentt}

The experiment was created with Google Forms.
However, Google Forms does not allow to upload files anonymously (it requires to login to upload, and the email of the uploader is included in the results).
As we wanted to avoid collecting any personally identifying information, we used the services of Formfacade, which integrates with Google Forms and allows to add various functionalities including anonymous file upload.
We also used Formfacade to create a website from the form, which was more aesthetic and convenient to distribute.
The form contained an introduction and 3 sections as follows.
\begin{enumerate}
    \item Initially we informed the participants that the experiment and the questions are not mandatory, they can quit the experiment whenever they want, and that no personal information is collected --- the experiment is completely anonymous.
    The introduction page also informed them that by moving to the next page they agree to participate under these conditions.
    \item After the introduction participants were asked questions related to their background, such as age, programming education, development experience, etc.
    \item For the experiment itself they got our flattened code as a link to GitHub, and were requested to download it.
    They could work on extracting the functions in their preferred IDE and work environment, with no time limit.
    After they finished refactoring it, they needed to make a zip file of their code and upload it.
    \item Finally, the participants were asked to complete a short survey with questions related to the code assignment they were requested to do.
    One question in the survey was about what they thought was the ideal length for functions.
    In addition we asked them to rate different considerations that may apply to function extraction refactoring.
\end{enumerate}

The link to the website we created was sent to personal contacts with familiarity with programming.
In addition, the experiment was published in two Reddit forums, r/SoftwareEngineering and r/Code.
We wanted the participants to be as varied as possible.

\begin{figure}\centering
\hspace*{-4mm}\includegraphics[width=1.05\columnwidth]{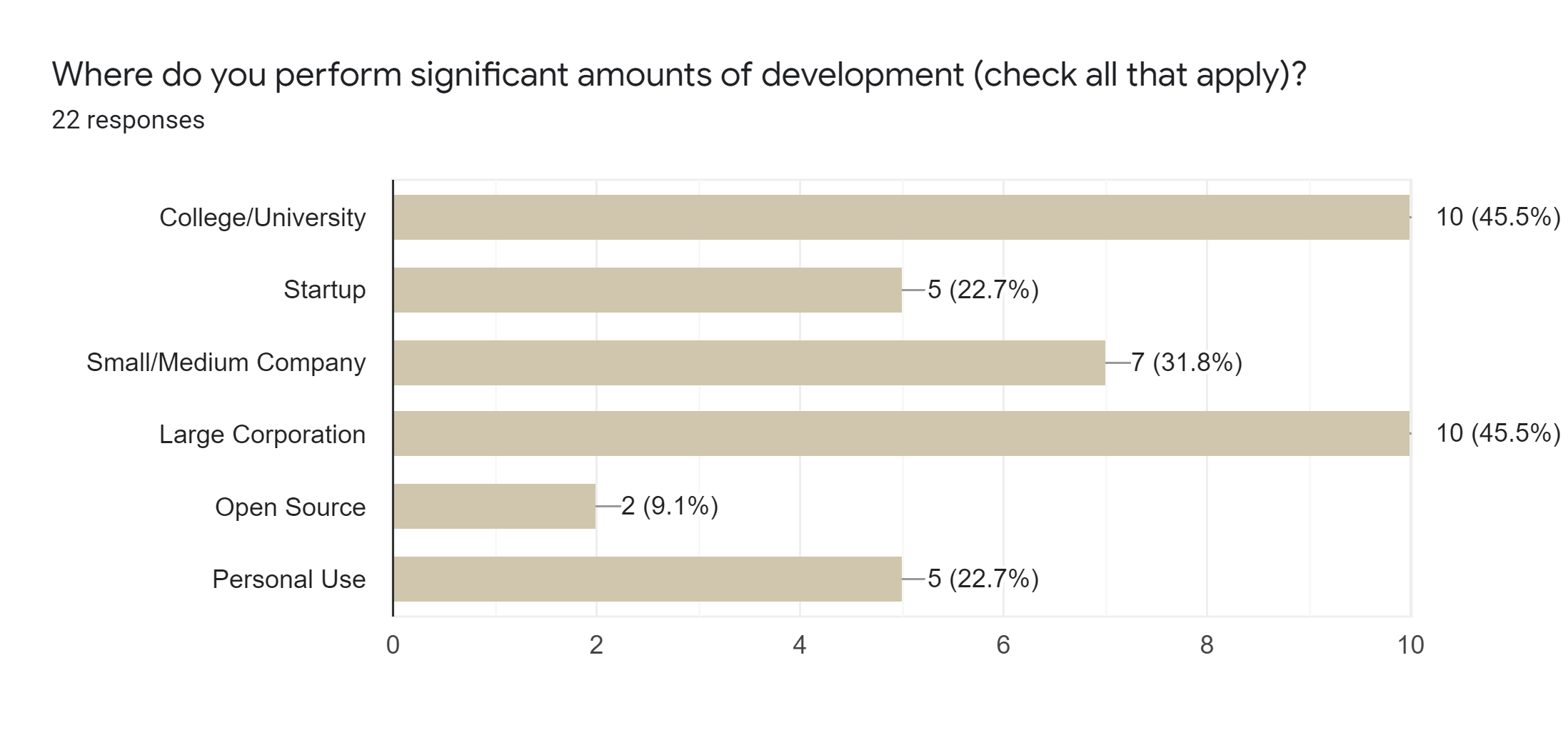}
\caption{\label{fig:dev_where}
Participants' development environments.}
\end{figure}

We received a total of 34 responses to the experiment.
Of these only 23 submitted the refactored code, which is the most important part; the 11 others were removed from the analysis.
The background of the participants was as follows.
Their ages were rather focused, with 72\% being between 28 and 30 years old; the oldest was 49.
51\% had a BSc/BA in a computer-related field, and another 30\% had an MSc; only 9\% were students who had not completed their first degree yet.
91\% reported that they were employed.
The range of development experience in industry, that is excluding studies, was from 1 to 29 years, but for 57\% it was up to 2 years.
48\% were employed as developers, 33\% had student jobs, and 10\% held management positions.
Most of the participants perform significant amounts of development in industry, in large corporations, small/ medium companies, or startups (in decreasing order --- see Figure \ref{fig:dev_where}).
Many also code in academic institutions or for personal use.
The most common programming language they felt comfortable with was Python (64\%), followed by Java (50\%) and C (36\%).
Finally, 44\% reported that they use agile development practices.

\subsection{Code Analysis}
\label{sect:analysis}

As noted above the code comprised 208 lines of actual code.
After refactoring it was a bit longer, due to the added function headers and function calls.
Analyzing this amount of code is hard and error prone.
We therefore wrote a set of scripts to analyze it.
Most of the analysis was done with regex matchers.

\subsubsection{Code Stripping}
\label{subsec:strip}

The first script strips the results from the irrelevant code parts such as whitespaces and blank lines.
The purpose of this script is to align all the results and make them comparable by changing only the way the code was presented, but not the content.
The script works in the following way:
\begin{enumerate}
    \item Remove all the import statements.
    \item Remove all the comment lines, and comment blocks demarcated with \code{"""} on the first and last line.
    \item Remove all blank lines.
    \item If a statement was divided into multiple lines, they were joined into one line.
    Specifically, if there is an opening bracket, \code{(}, but no closing one, \code{)}, all the following lines were concatenated to the initial line until the line with the closing bracket.
    \item Remove all the white space at the beginning and at the end of each line.
    Since Python is a language where indentation is part of the syntax, this step must be done last.
    Its purpose is to make the comparison between the lines independent of the exact indentation used.
\end{enumerate}

\subsubsection{Lines Identification}
\label{subsec:lines}

The code submitted by the experiment participants needs to be compared to the original code.
This comparison was based on comparing individual lines.
For instance, to identify which part of the participant's code was originally in a function, we needed to compare each line in the participant's code to the original code.
At first, we compared between the lines just by using the Python "==" operator after we cleaned the white spaces, but it was not good enough.
For example, where were many false negatives where the participants changed a variable name, changed the argument name the function receives, or just added an underscore.

Because of these problems, we switched to comparing with \code{SequenceMatcher} from the \code{difflib} Python library.
This provides a similarity ratio between 0 and 1.
We could then choose a threshold to identify lines with small changes as equal, and lines with more changes as different.
The threshold we used for comparing lines was 0.95.
The threshold for finding lines that were in the original code but were not in the results (lines that were not kept) was 0.75 (these are the lines denoted by black dots in Figure \ref{fig:res-flat} below).

\subsubsection{Functions Identification}
\label{subsec:functions}

To compare code refactoring by extracting functions we need to identify the functions in the results.
This was done by a script which counts the indentation and looks for the \code{def} keyword which defines the start of a function. 
When a \code{def} is found, the string after it up to the character \code{(} is taken as the name of the function.
The body of the function is all following lines with a larger indentation.

Once all the lines in a function have been identified, we retrieve from the original code the range of lines that the function was extracted from.
First the line which starts the function is checked for its line number in the original code.
If the line does not appear in the original code, the next line is checked.
This process continues until a line appears in the original code. 
After that, the line which ends the function is checked for the line number in the original code using the same process.
If the line does not appear in the original code, the previous line is checked until a line appears in the original code.
The range is then determined by these lines, and we verified that it matches the length of the function that was extracted.
Overall there were 61 unidentified lines in a total of 142 extracted functions.
Most of these (57) were the \code{return} lines which ended these functions.
Obviously, these lines did not appear in the flattened code, because the flattened code contained only one main function.
The remaining 4 lines were not identified due to extreme changes that were made by the participants, or were completely new lines they decided to add.

We also checked if the lines' order was changed in relation to the original code.
This is done by running over all the functions and for each function checking if there are 2 lines which appear in different order than they appear in the original code.
This is done by looking at every pair of successive lines, retrieving their lines numbers in the original code, and validating that those numbers are also in the same order relation as the lines in the function.
In all the results we had only 1 case where the order of 2 lines was changed.

\subsubsection{Manual Analysis}
\label{subsec:manual}

A manual scanning of the results was also done.
We did this for two main purposes:
\begin{enumerate}
    \item Verify the scripts: we wanted to do some manual testing to make sure the scripts are accurate.
    \item Looking for patterns that can't be seen in the scripts: We tried to retrieve more information about the function extraction by understanding what the participants' main considerations for the refactoring were.  
\end{enumerate}

\section{Results and Discussion}
\label{sect:Results}

\begin{figure*}\centering
\includegraphics[width=1.3\columnwidth,trim=11mm 0mm 5mm 20mm,clip]{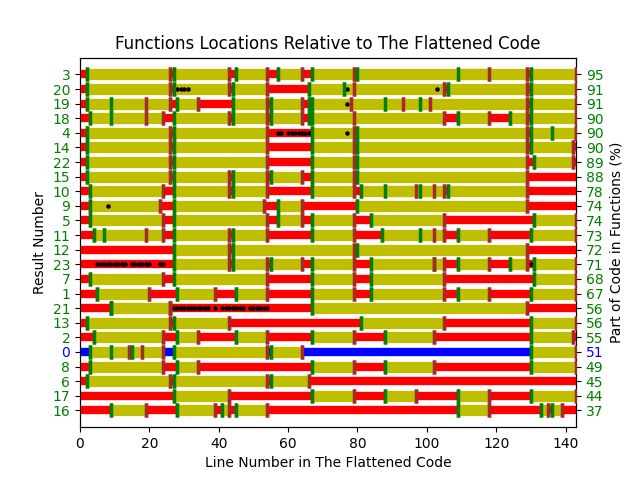}
\caption{Visualization of extracted functions locations relative to the flattened code.
The horizontal dimension represents location in the flattened code, with the yellow-olive segments showing lines of code that were extracted to functions.
Green and brown lines indicate function start and end, respectively.
Each horizontal bar represents the refactoring performed by one participant.
Result 0 (in blue) is the original code.}
\label{fig:res-flat}
\end{figure*}

\subsection{Function Extraction Results}

The participants in the study were requested to provide us with two inputs: the refactored code, and answers to a set of questions.
We begin with the code analysis.

We analyzed the coding task results both manually and automatically with scripts as described above.
Figure \ref{fig:res-flat} shows a bird's-eye view of all the function extractions by all the participants, compared with the original code.
The horizontal dimension represents the lines of code in the original flattened code, after excluding all the blank and comment lines.
The range is from the first line at the left to the last line, line 143, at the right.
This is less than the original 208 lines of actual code because in the original layout arguments to functions were spread over multiple lines, one each, for readability.
As we noted above, in the analysis we unified such argument lists in a single line.

Each horizontal bar represents one experimental subject.
The red segments are lines of code that were left in the main function.
The yellow-olive parts are ranges that were extracted to functions.
The numbers on the left are the participants' serial numbers in the results data.
They are sorted by the fraction of the code that they extracted into functions, as noted on the right-hand side.
The blue line, with serial number 0, is the original code.
The black dots represent lines in the results that did not appear in the original flattened code, meaning these lines were replaced by new lines or were changed too drastically to allow the comparator to find them in the original code.

\begin{figure}\centering
\includegraphics[width=1.08\columnwidth,trim=11mm 0mm 5mm 20mm,clip]{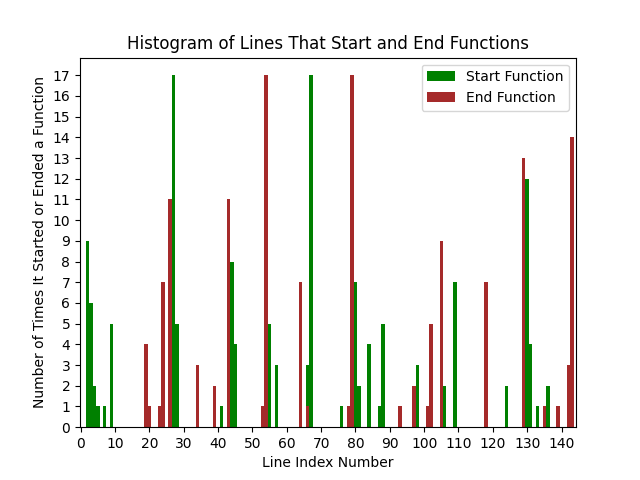}
\caption{Lines that start and end functions.}
\label{fig:func-start-end}
\end{figure}

Looking at the figure, it is immediately obvious that function extraction is often correlated across many participants.
Starting to analyze this data, we find that all the participants extracted between 3 and 10 functions, with just over half extracting 5--7 functions.
Figure \ref{fig:func-start-end} shows where in the flattened code functions start and end.
This was calculated by taking the flattened code as a reference, that is the number of each line that was identified as starting or ending a function was determined by looking for its line number in the flattened code.
If a line was not found in the flattened code, it was ignored, and the closest line that was found was chosen, as described above in Section \ref{subsec:functions}.
The graph shows that there are many cases where the end of one function is closely followed by the beginning of another.
But in many other cases functions are separated by blocks of code that remained in the main function.
Notably the line at the beginning of the whole code was always left out, which is expected: it is the \code{def} of the main function.

\begin{figure}\centering
\includegraphics[width=1.08\columnwidth,trim=11mm 0mm 5mm 20mm,clip]{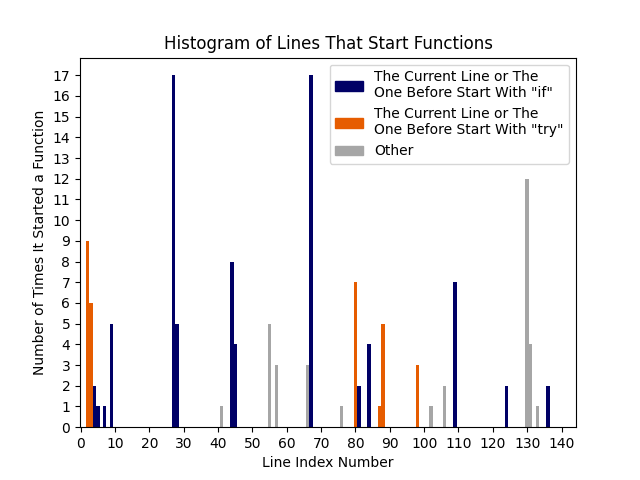}
\caption{Lines that start functions by type.}
\label{fig:func-start}
\end{figure}

The lines that start functions are shown again in Figure \ref{fig:func-start}, but here they are colored by the code which appears in that line (or adjacent to it): \code{if} in dark blue, \code{try} in orange, and other in gray.
As we can see the vast majority of functions start on an \code{if} (77 of the 142 functions defined by the experiment participants) or \code{try} (31 functions).
In addition, near the end of the code there is a block where the response object is created, and many of the participants extracted this block of code to a function.
These are the functions that start on line 130 (including the creation of the object itself) or line 131 (only the population of the object's data members).

Our conclusion is that the main triggers for extracting functions in our experiment are coding constructs.
This could be related to the Python syntax, which requires such constructs to be indented.
The start of a new block of indented code may then be taken as an indication to extract a function.
Comparing with the data about function ends from the previous figure, we find that the main trigger for ending functions was the same --– namely the end of an \code{if} or \code{try} block.
Notably, our results resonate with the recommendations made by Martin in \textit{Clean Code} \cite{martin:clean}, that try-catch blocks be extracted and that the body of an \code{if} should be one line long and specifically a function call.

\begin{figure}\centering
\includegraphics[width=1.08\columnwidth,trim=11mm 0mm 5mm 20mm,clip]{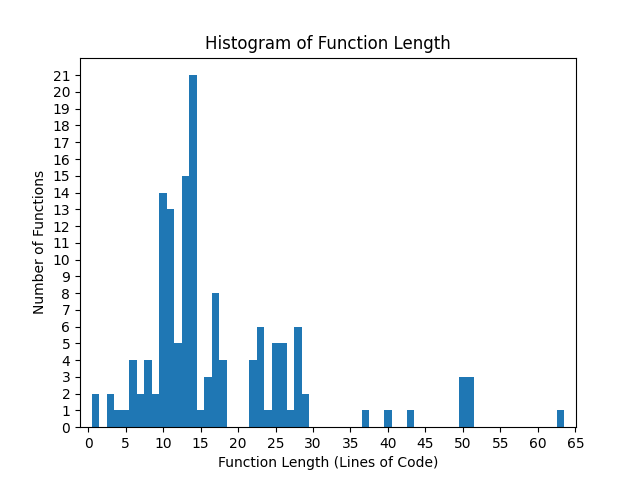}
\caption{Histogram of function lengths.}
\label{fig:func-len}
\end{figure}

Figure \ref{fig:func-len} shows the lengths of the functions the participants chose to extract from the flattened code (all the functions extracted by all the participants, but not the main function).
As we can see in the figure, 80\% of the extracted functions had 5--30 lines of code, and 55\% of the functions had a length of 10--14 lines.
This result is discussed below in connection to the survey question about the ideal function length.

\begin{figure}\centering
\includegraphics[width=1.05\columnwidth,trim=9mm 0mm 5mm 20mm,clip]{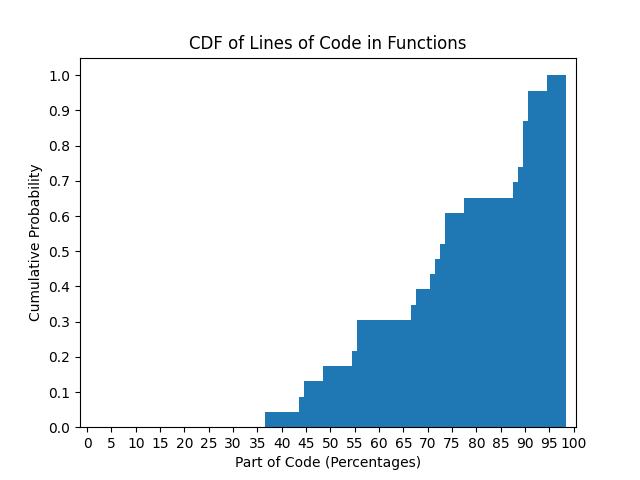}
\caption{CDF of lines of code in functions.}
\label{fig:code-in-func}
\end{figure}

Figure \ref{fig:code-in-func} shows the part of code in percents that was extracted to functions.
This is a cumulative distribution function (CDF).
The horizontal axis is the percentage of lines that where extracted.
The graph shows, for each percentage, what fraction of the participants had extracted that percentage of lines or less.

As we can see the majority of the participants extracted at least 70\% of the code to functions.
The reason may be a general understanding that code should be divided into functions and not be structured as one big main function.
But on the other hand there might be a bias due to the definition of the experiment.
The experiment instructions were to extract functions, and this might cause the participants to perform more function extractions than they would otherwise.
In addition, if we compare the results to the original code, the original code had only 51\% of the code in functions, placing it fifth from the end in this parameter.
This reinforces the conjecture that there may be a bias because of the experiment definition.

A very important observation from Figure \ref{fig:code-in-func} is the diversity in the fraction of code extracted to functions.
This ranges from a low value of extracting only 37\% of the code lines, leaving nearly two thirds of the code in the main function, to a high mark of extracting no less than 95\% of the code, leaving nearly nothing in the main function.
And between these extremes we see a wide distribution of values.
This testifies to differences of opinion between the experiment participants regarding what parts of the code should be extracted to functions.

To get a better picture of these differences, we can return to Figure \ref{fig:res-flat}.
This figure enables us to identify 6 blocks of code that were the main candidates for extraction:
\begin{enumerate}
    \item Lines 3 to 26, extracted (at least partially) by 20 participants and in the original code.
    This is a big \code{try} block surrounding the creation of the connection for sending the request.
    \item Lines 27 to 54, extracted (at least partially) by 22 participants and in the original code.
    This verifies the SSL certificate, and contains two big \code{if} blocks, which were extracted together or separately.
    \item Lines 55 to 64, extracted by only 9 participants, but also in the original code.
    This requests the URL and contains a smaller \code{if}.
    \item Lines 65 to 79, extracted by 19 participants, but not in the original code.
    This is a large \code{if} block that checks whether a timeout is needed.
    It was nearly always extracted in exactly the same way.
    \item\label{item:try} Lines 80 to 129, extracted (at least partially) by 22 participants but not in the original code.
    This large block of code
    is where the actual sending is done and the response received.
    But these operations may fail for myriad reasons, so they are surrounded by three nested \code{try-except} blocks and error handling code.
    There was especially high variability concerning what parts of this to extract, reflecting different opinions regarding whether to extract the full largest \code{try-except} block as one large function, or to extract only some internal parts from it.
    \item Lines 130 to 143, extracted by 17 participants and in the original code.
    This is code that creates the object to return, and it was nearly always extracted in exactly the same way.
\end{enumerate}
Interestingly, three of these code blocks (numbers 1, 2, and 6) correspond to extraction types identified by Hora and Robbes as the most common, which are for creation, validation, and setup \cite{hora20}.

To summarize, the variability in the amount of code extracted to functions stems from different opinions on whether certain blocks should be extracted, whether to include or exclude nested constructs, and whether to include or exclude a surrounding construct (e.g.\ include the \code{if} itself or only its body).

Note that part of the diversity concerns \code{try-except} blocks and error handling code.
This reflects the fact that indeed there are different ways and designs to handle errors.
Some believe that error handling logic should be located in conjunction with the functions which suffered the errors, while others believe that it should be separated such that all errors are handled together.
It is worthwhile to mention that in the original code this error handling code was not extracted to a dedicated function.

\subsection{Survey Results}
\label{sec:codeing-survey} 

After the code assignment the participants were asked questions about the considerations involved in function extraction.
We started with asking them about the ``clean code'' approach to code development.
61\% of the participants indicated that they were familiar with this approach.
And of those that were familiar, 93\% said they agreed with the principles of clean code and strive to act on it (answers of 4 or 5 on a scale of 1 to 5).

\begin{figure}\centering
\includegraphics[width=1.08\columnwidth,trim=11mm 0mm 5mm 20mm,clip]{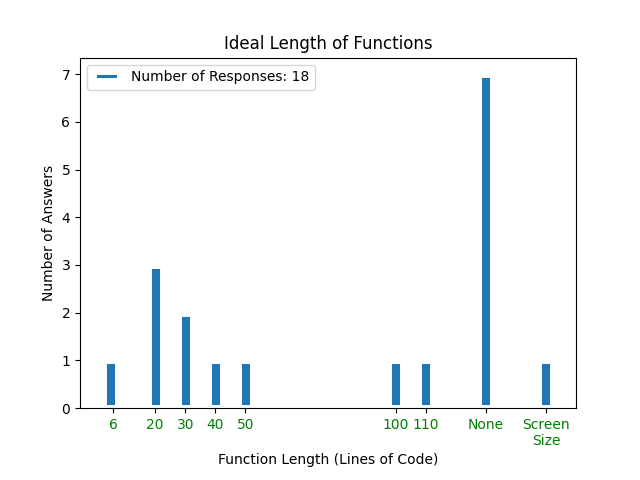}
\caption{Responses to the question concerning the ideal length of functions.}
\label{fig:ideal-func-len}
\end{figure}

The next question concerned the ideal function length.
The responses we received are shown in Figure \ref{fig:ideal-func-len}.
As we can see the most common answer was that there is no ideal length of function, but from those who did respond most of the responses were 20--30 lines of code.
Comparing this to the results of the actual functions they extracted, shown above in Figure \ref{fig:func-len}, we find that the actual functions were even shorter than the ideal length the participants stated they believe in.
Interestingly, the actual results agree with the recommendation that functions should usually have no more than 20 lines, suggested by Martin in \textit{Clean Code} \cite[p.~34]{martin:clean}.

\begin{table}\centering
\caption{Main considerations for function extraction}
\label{table:considerations}
\begin{tabular}{|p{35mm}|c|c|}
\hline
 \emph{Consideration} & \emph{Mean} & \emph{Histogram} \\
\hline\raggedright
 Making each function's logic cohesive\rule[-8mm]{0mm}{8mm} & 4.62 & 
\raisebox{-10mm}{\includegraphics[scale=0.125,trim=8mm 5mm 15mm 10mm,clip]{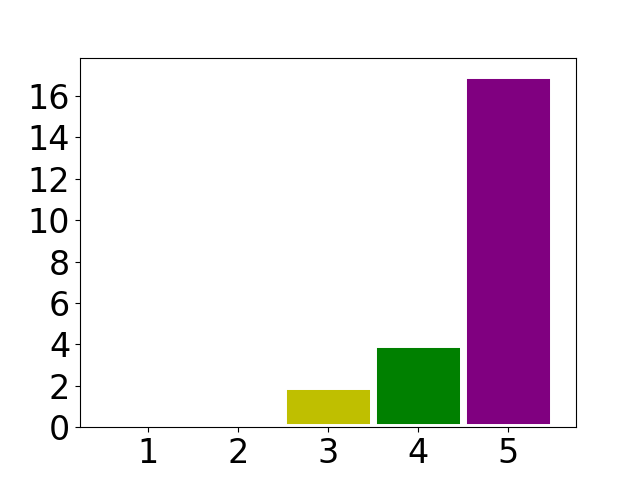}} \\
\hline\raggedright
 Making each function do only one thing\rule[-8mm]{0mm}{8mm} & 4.10 & 
\raisebox{-10mm}{\includegraphics[scale=0.125,trim=8mm 5mm 15mm 10mm,clip]{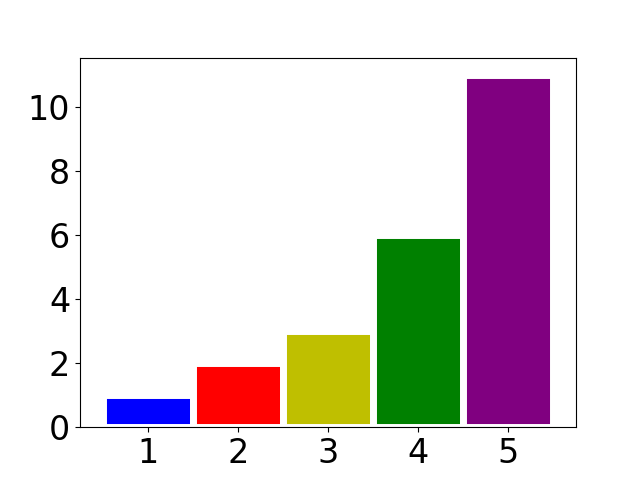}} \\
\hline\raggedright
 Separating code control blocks\rule[-10mm]{0mm}{10mm} & 3.93 & 
\raisebox{-9mm}{\includegraphics[scale=0.125,trim=8mm 5mm 15mm 10mm,clip]{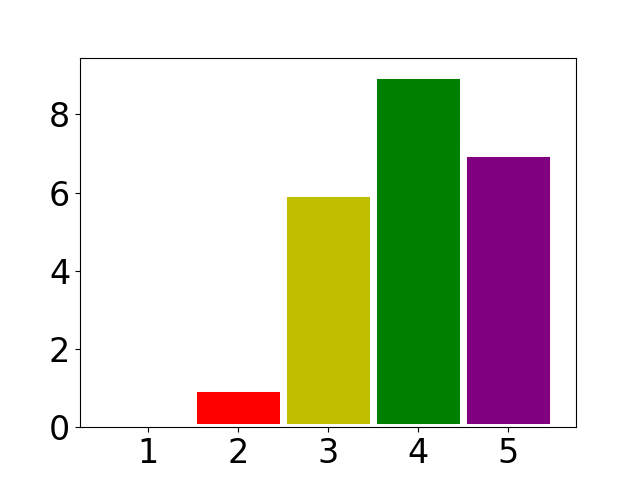}} \\
\hline\raggedright
 Making functions easily testable\rule[-8mm]{0mm}{8mm} & 3.89 & 
\raisebox{-10mm}{\includegraphics[scale=0.125,trim=8mm 5mm 15mm 10mm,clip]{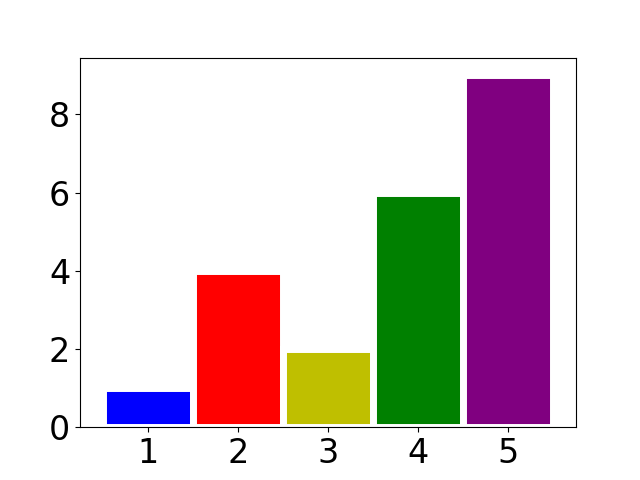}} \\
\hline\raggedright
 Following design patterns\rule[-10mm]{0mm}{8mm} & 3.41 & 
\raisebox{-9mm}{\includegraphics[scale=0.125,trim=8mm 5mm 15mm 10mm,clip]{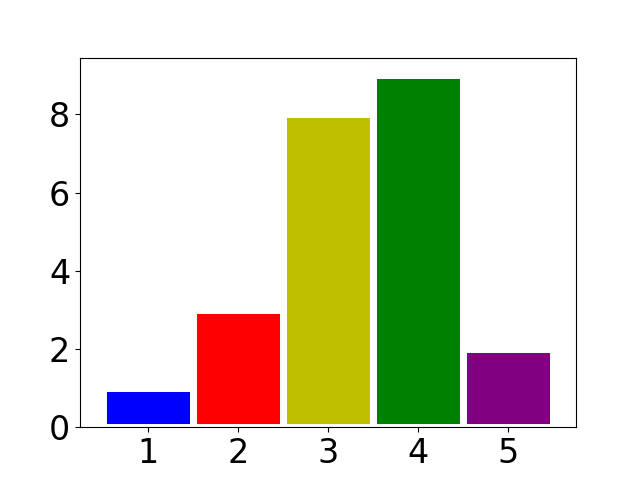}} \\
\hline\raggedright
 Not needing to pass too many arguments\rule[-8mm]{0mm}{8mm} & 3.10 & 
\raisebox{-10mm}{\includegraphics[scale=0.125,trim=8mm 5mm 15mm 10mm,clip]{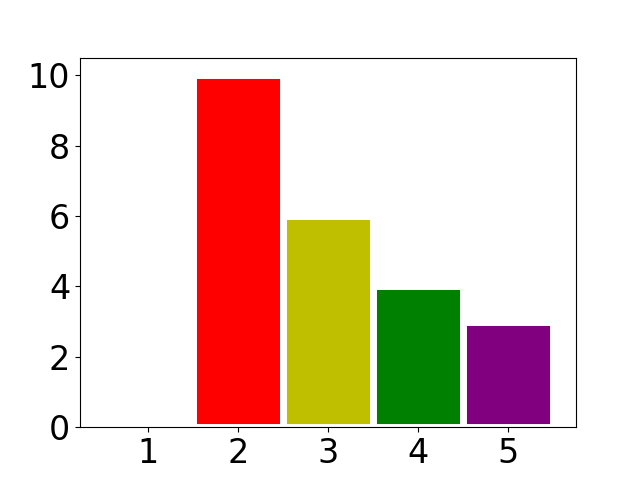}} \\
\hline\raggedright
 Making functions as short as possible\rule[-8mm]{0mm}{8mm} & 2.72 & 
\raisebox{-10mm}{\includegraphics[scale=0.125,trim=8mm 5mm 15mm 10mm,clip]{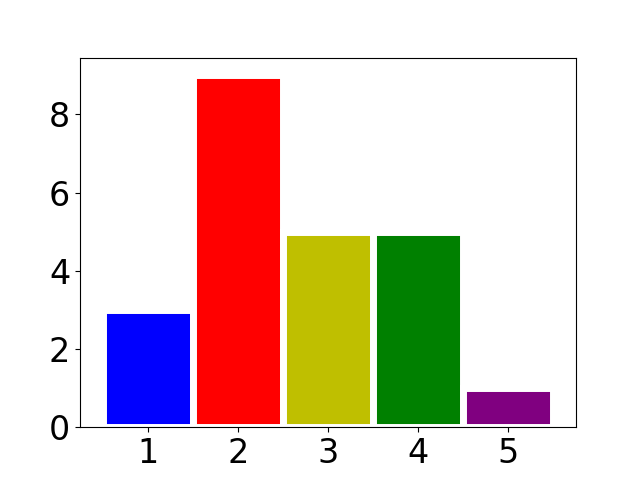}} \\
\hline\raggedright
 Making each function close to the ideal length\rule[-8mm]{0mm}{8mm} & 2.65 & 
\raisebox{-10mm}{\includegraphics[scale=0.125,trim=8mm 5mm 15mm 10mm,clip]{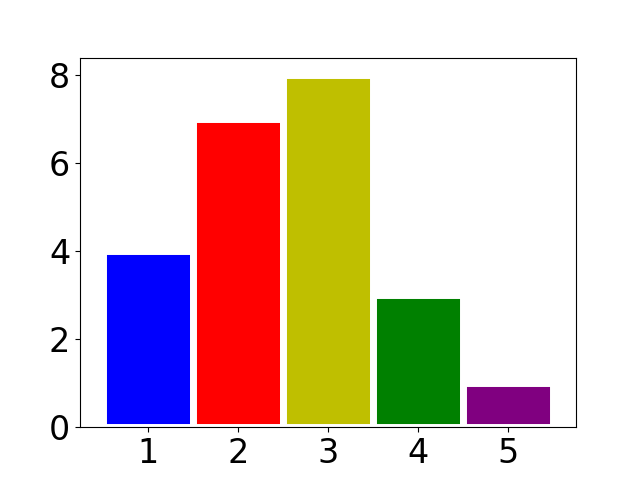}} \\
\hline
\end{tabular}
\end{table}

Finally, we asked about the participants' main considerations when extracting new functions.
The distribution of answers for the different considerations we suggested are shown in Table \ref{table:considerations}.
The scale was from 1 = Not important to 5 = Very important.
As we can see, the most popular consideration was ``Making each function's logic cohesive'' with an average score of 4.62/5.
The consideration that corresponds to our observation in Figure \ref{fig:func-start} is the third most popular consideration --- ``Separating code control blocks'', with an average score of 3.93/5.
We can't really measure the actual difference in the importance between these two considerations by looking at the function extraction results, because ``Making each function's logic cohesive'' is quite an abstract consideration, and incomparable to the ``Separating code control blocks'' consideration which can be measured by counting the lines that started a function on a new block of code.
Interestingly, the issue of function length was ranked as having lower importance than all other considerations.

\subsection{Discussion}
\label{sec:discuss} 

One of our main results is that participants in the experiment tend to extract functions based on the blocks of code defined by existing coding constructs, such as \code{if} and \code{try}.
They also gave the consideration of ``Separating code control blocks'' relatively high scores in the survey.
Using blocks of code as cues for function extraction is reasonable because such blocks indeed encapsulate separate sections of the computation.
However, we note that extracting functions by looking at blocks of code is also the easiest way to extract.
It doesn't even require the participants to understand the code's logic.

Another interesting observation is the difference between the approaches taken by tools and by human developers.
As we noted above in the related work section, tools are often based on extracting program slices (e.g.\ \cite{maruyama01,tsantalis11}).
Program slices are defined by the data flow in the program: they include instructions that may affect the value of a certain variable.
This is useful for code analysis, and to ensure that extracting the code retains exactly the same behavior.
But following data flow is not easy for humans.
Therefore the participants in our experiment preferred to extract functions based on coding constructs, which reflect control flow.
It may be interesting to contrast these approaches in terms of the similarities and differences between the refactored codes they produce.
We leave such a study to future work.
In the meanwhile, we note that our results resonate with those of Pennington from 25 years ago, who found that professional programmers base their mental representations of programs on control flow, and not on functionality \cite{pennington87}.

Another interesting issue is the question of ideal function length.
As we noted, Martin famously makes the case that functions should be as short as possible \cite[p.~34]{martin:clean}.
Given that 61\% of our participants acknowledged that they are familiar with the clean code discipline, this might be the reason that 80\% of the functions were 10-30 lines long, and 55\% had 10-14 lines.
However, in the survey we saw that ``Making functions as short as possible'' and ``Making each function close to the ideal length'' were the two least important considerations for function extraction (Table \ref{table:considerations}).
Thus the participants believe that small functions and ideal functions length are not as important as at least 6 other considerations we suggested to them.
But in fact they do extract small functions.
This is quite an interesting result, because it means the participants extracted small functions without explicitly thinking about then in this way.
We might carefully conjecture that nowadays developers extract small function as part of their coding skills.
Another possibility is that other considerations, like keeping the logic cohesive and doing just one thing, naturally lead to shorter functions.
So short functions are not an end in themselves, but they happen to be the solution to other goals.

\section{Threats to validity}

As in any experiment, we had some difficulties and threats to the validity of our results.
The following threats are the most significant remaining ones.

\subsection{Confounding factors}

Our experiment was explicitly about extracting functions.
This may have affected the participants, who may have tried to ``live up to expectations'' and extracted more than they would have under real work conditions.
Such behavior, if it exists, would lead to a bias in the results.

Another possible effect of an experiment environment is that it is perceived as less serious than ``real'' work, leading participants to invest less effort.
This was noted in a few comments we received to the postings inviting participants on reddit.
One wrote ``Living organisms are lazy (including humans) and will usually choose the cheapest way to get out of a problem. So your experiment might tell you what's the cheapest way to reformat code''.
Another wrote ``this is a 150 lines function, there were occasions where reformatting something like that was a task for the whole day in my job''.
Strengthening this, at the end of the survey we asked whether seeing the survey questions might have changed the way the code was refactored.
48\% answered yes, possibly implying that they did not think about what they had done deeply enough.

It is not clear how these problems could be avoided in an experiment.
A possible approach is to conduct a much larger experiment, in both scope and cost, in which developers are actually hired to generally refactor a large body of code \cite{sjoberg02}.
Such as experiment can be attempted based on our experiment's demonstration that the methodology of using flattened production code looks promising.

\subsection{Limited generalizability}

Being an initial experiment in a new direction, and given the need to develop a new methodology, our experiment was rather limited.
We used only one function, which raises the question of how representative this is of other codes.
For example, one of the main reasons for creating functions is to reuse code, so if the code contains clones they may be prime targets for extracting.
However, our code did not contain any clones, so this was not checked.
In addition, our results are limited to the context of refactoring, and may not generalize to function creation in the context of writing new code.

We also had only 23 participants who submitted the refactored code.
Given the need to invest around half an hour in this non-trivial task this is not a bad response.
However, it is not enough to enable an analysis of the behavior of different demographic groups, as each one would have too few members.
Another issue is whether our participants are representative of developers in general.
For example, in our experiment only 61\% of the participants indicated they know clean code, while in a survey by Ljung et al.\ the vast majority of the participants had heard of clean code, and tended to agree with its principles \cite{ljung22}.
Again, based on our initial experiment, a larger one can be attempted.

\section{Conclusions}
\label{sect:Conclusions}

\subsection{Summary of Results}

In this study we tried to make initial observations of how developers create functions, including issues like how much they agree with each other and how long are the functions they create.
Our approach was to design an experiment, where we took real production code, ``flattened'' it, and then asked the experiment participants to refactor it by extracting functions.
In addition, we asked them to answer some questions related to their considerations when extracting the functions.

Analyzing the results we found that most of the participants extracted small functions, as suggested in the ``Clean Code'' approach, and that the main cues for extracting functions were related to the code structure.
Interestingly the actual functions were somewhat shorted than what the participants said would be ideal.
Also, they didn't explicitly consider length to be a major factor, citing logical cohesion and doing only one thing as the most important considerations.

We also found that in most cases the fraction of the code that was extracted to functions was bigger than it had been in the original code.
This was probably due to the fact we asked participants explicitly to extract functions.
But it also implies that developers can be influenced in how they go about designing functions.
This can be studied in future experiments. 

We believe that in addition to these concrete results, we produced some promising methods to study such complicated issues like software design.
Since this kind of experiments are almost non-existent, we hope our study and our methods will encourage additional experiments in this field.

\subsection{Implications}

Our work has several possible implications for practitioners.
The most immediate is to highlight a simple and accessible approach to function extraction.
Extracting functions is a natural and intuitive type of refactoring, which perhaps explains why it is one of the most commonly used types of refactoring.
And by focusing on code blocks defined by constructs like loops, conditionals, or exception handling one can perform function extraction in a straightforward manner.

Extracting such blocks also provides a simple methodology to control the desired granularity, as one can continue to extract nested blocks down to the most basic blocks, or alternatively stop at a larger scope.
For example, when extracting an \code{if} construct to a separate function, one can then decide whether to also extract the ``then'' and ``else'' blocks or not.
If one desires to create very short functions, extracting blocks is an easy way to achieve this.

Using code blocks for function extraction is also related to program comprehension.
When faced with unknown code, understanding what it does requires a bottom-up approach \cite{shneiderman79}.
This is greatly simplified if each function is short, and composed mainly of calls to other functions which have good descriptive names, as such a structure allows the readers to perceive the code at a higher level of abstraction rather than forcing them to contend with the basic constructs.
Extracting blocks achieves exactly this effect.

\subsection{Future Work}
\label{sect:future}

We have reported on an initial experiment on function extraction, based on a single code base.
An obvious line of future work is therefore to explore the wider validity of our results, by replicating the experiment under diverse conditions.
In particular, it is important to create and run experiments with more code, and to compensate developers for investing time to refactor it as they would in actual work conditions.
Will control structures remain the main cues for function extraction under such conditions, or will other considerations emerge?

Another line of further research is to elaborate on our results.
One of our important observations is that there is no unanimous agreement on exactly how to extract functions.
But each such difference of opinion is actually a research question in disguise.
For example, one of our main results was that control structures such as \code{if} serve as scaffolding for function extraction.
But our study participants differed in their opinions whether the \code{if} instruction itself should be part of the extracted function, or perhaps only the body of the \code{if} should be extracted.
A related issue concerns scope: should every \code{if} be extracted, or only \code{if}s above a certain size?
What are the implications in terms of code comprehension and maintenance?
Answering such questions may provide very useful guidelines for practitioners facing low-level design decisions.

Our experiment on function extraction is a suitable starting point for empirical research on design since functions are the most basic elements of code design.
However, the core of software design is concerned with modularity \cite{parnas72}, and especially with the definition of classes and the decision of what functionality should be included in each class.
One can envision an experiment analogous to ours, where one takes a software package comprising multiple classes, ``flattens'' it to form a large monolithic program, and then asks participants in the experiment to refactor it and extract classes and methods.

We have tried to design such an experiment and found that it is much harder to design than the current experiment on functions.
The code we used was an academic project in Java implementing a simplified banking operation.
This code included 5 classes: \code{Bank}, \code{Account}, \code{User}, \code{Transaction}, and \code{ATM}.
The focus was on the operation of the ATM.
The problem with transforming this into a monolithic program was that classes contain state.
At runtime multiple objects are created to represent the actors and operations involved in the ATM's operation, and the data about each actor and operation is maintained in the corresponding object.
But when we create a monolithic program there is no longer any natural place to maintain this data.
We need some new global data structures or an external database.

To test this experiment design we created a pilot in which the data was stored in an array of arrays, and asked 4 subjects to refactor it.
The results were disappointing for two reasons.
First, it took too much time and frustrated the subjects.
But worse, the structure of the arrays containing the data dictated the way that they partitioned the program into classes.
This undermined the whole rationale of the experiment, and implies that another approach is needed.
We therefore leave the design of controlled experiments on class extraction to future work.

\section*{Experimental Materials}

Experimental materials are available at
\begin{center}
    https://doi.org/10.5281/zenodo.7044101
\end{center}


\bibliographystyle{myabbrv}
\bibliography{abbrv,se}

\end{document}